\documentclass[twocolumn,times,tighten,Astrosymb]{aastex631}
\usepackage{xcolor}
\newcommand{\ovi}{\ion{O}{6}}
\newcommand{\cii}{\ion{C}{2}}
\newcommand{\siii}{\ion{Si}{2}}

\newcommand{\nv}{\ion{N}{5}}

\newcommand{\msun}{$\rm{M_{\odot}}$}

\usepackage{multirow}
\usepackage{amsmath}
\usepackage{lipsum}
\newcommand{\angstrom}{\textup{\AA}}
\newcommand{\Ub}{\emph{U}-band }%
\newcommand{\sextractor}{\textsc{SExtractor} }
\newcommand{\swarp}{\textsc{swarp} }
\revised{\today}
\accepted{\today}
\submitjournal{PASP}

\shorttitle{Searching for Diffuse Intragroup UV Light}
\shortauthors{McCabe et al.}

\begin{document}

\title{Searching for Intragroup Light in Deep U-band Imaging of the COSMOS Field}

\correspondingauthor{Tyler McCabe}
\email{tyler.mccabe@asu.edu}

\author[0000-0002-5506-3880]{Tyler McCabe}
\affiliation{School of Earth and Space Exploration, Arizona State University, Tempe, AZ 85287-1404, USA}
\author[0000-0002-9961-2984]{Caleb Redshaw}
\affiliation{School of Earth and Space Exploration, Arizona State University, Tempe, AZ 85287-1404, USA}
\author{Lillian Otteson}
\affiliation{School of Earth and Space Exploration, Arizona State University, Tempe, AZ 85287-1404, USA}
\author[0000-0001-8156-6281]{Rogier A. Windhorst}
\affiliation{School of Earth and Space Exploration, Arizona State University, Tempe, AZ 85287-1404, USA}
\author[0000-0003-1268-5230]{Rolf A. Jansen}
\affiliation{School of Earth and Space Exploration, Arizona State University, Tempe, AZ 85287-1404, USA}
\author[0000-0003-3329-1337]{Seth H. Cohen}
\affiliation{School of Earth and Space Exploration, Arizona State University, Tempe, AZ 85287-1404, USA}
\author[0000-0001-6650-2853]{Timothy Carleton}
\affiliation{School of Earth and Space Exploration, Arizona State University, Tempe, AZ 85287-1404, USA}
\author[0000-0002-2724-8298]{Sanchayeeta Borthakur}
\affiliation{School of Earth and Space Exploration, Arizona State University, Tempe, AZ 85287-1404, USA}
\author[0000-0003-4439-6003]{Teresa A. Ashcraft}
\affiliation{School of Earth and Space Exploration, Arizona State University, Tempe, AZ 85287-1404, USA}

\author[0000-0002-6610-2048]{Anton M. Koekemoer}
\affiliation{Space Telescope Science Institute, Baltimore, MD 21218, USA}

\author[0000-0003-0894-1588]{Russell E. Ryan}
\affiliation{Space Telescope Science Institute, Baltimore, MD 21218, USA}

\author[0000-0001-6342-9662]{Mario Nonino}
\affiliation{INAF - Osservatorio Astronomico di Trieste, Via Bazzoni 2, 34124 Trieste, Italy}
\author[0000-0002-7409-8114]{Diego Paris}
\affiliation{INAF - Osservatorio Astronomico di Roma, Via Frascati 33, I-00078 Monte Porzio Catone, Italy}
\author[0000-0002-5688-0663]{Andrea Grazian}
\affiliation{INAF - Osservatorio Astronomico di Padova Vicolo dell'Osservatorio, 5 Padova (PD) I-35122, Italy}
\author[0000-0003-3820-2823]{Adriano Fontana} 
\affiliation{INAF - Osservatorio Astronomico di Roma, Via Frascati 33, I-00078 Monte Porzio Catone, Italy}
\author[0000-0003-0734-1273]{Emanuele Giallongo}
\affiliation{INAF - Osservatorio Astronomico di Roma, Via Frascati 33, I-00078 Monte Porzio Catone, Italy}
\author[0000-0003-3754-387X]{Roberto Speziali} 
\affiliation{INAF - Osservatorio Astronomico di Roma, Via Frascati 33, I-00078 Monte Porzio Catone, Italy}
\author[0000-0003-1033-1340]{Vincenzo Testa}
\affiliation{INAF - Osservatorio Astronomico di Roma, Via Frascati 33, I-00078 Monte Porzio Catone, Italy}
 
\author[0000-0003-4432-5037]{Konstantina Boutsia}
\affiliation{Carnegie Observatories, Las Campanas Observatory, Colina El Pino, Casilla 601, La Serena, Chile}

\author[0000-0002-8190-7573]{Robert W. O'Connell} 
\affiliation{Department of Astronomy, University of Virginia, Charlottesville, VA 22904-4325, USA}

\author[0000-0001-7016-5220]{Michael J. Rutkowski} 
\affiliation{Department of Physics \& Astronomy, Minnesota State University, Mankato, Mankato, MN 56001}

\author[0000-0002-9136-8876]{Claudia Scarlata}
\affiliation{Minnesota Institute for Astrophysics, University of Minnesota, 116 Church Street SE, Minneapolis, MN 55455, USA}

\author[0000-0002-7064-5424]{Harry I. Teplitz}
\affiliation{Infrared Processing and Analysis Center, MS 100-22, Caltech, Pasadena, CA 91125, USA}

\author[0000-0002-9373-3865]{Xin Wang}
\affiliation{Infrared Processing and Analysis Center, MS 100-22, Caltech, Pasadena, CA 91125, USA}

\author[0000-0002-9946-4731]{Marc Rafelski}
\affiliation{Space Telescope Science Institute, Baltimore, MD 21218, USA}
\affiliation{Department of Physics \& Astronomy, Johns Hopkins University, Baltimore, MD 21218, USA}
 
\author[0000-0001-9440-8872]{Norman A. Grogin}
\affiliation{Space Telescope Science Institute, Baltimore, MD 21218, USA}

\author[0000-0003-1581-7825]{Ray A. Lucas}
\affiliation{Space Telescope Science Institute, Baltimore, MD 21218, USA}

\begin{abstract}
We present the results of deep, ground based \Ub imaging with the Large Binocular Telescope of the Cosmic Evolution Survey (COSMOS) field as part of the near-UV imaging program, UVCANDELS. We utilize a seeing sorted stacking method along with night-to-night relative transparency corrections to create optimal depth and optimal resolution mosaics in the $U$-band, which are capable of reaching point source magnitudes of AB$\sim$26.5 mag at 3$\sigma$. These ground-based mosaics bridge the wavelength gap between the \textit{HST} WFC3 F275W and ACS F435W images and are necessary to understand galaxy assembly in the last 9-10\,Gyr. We use the depth of these mosaics to search for the presence of $U$-band intragroup light (IGrL) beyond the local Universe. Regardless of how groups are scaled and stacked, we do not detect any $U$-band IGrL to unprecedented \Ub depths of $\sim$29.1--29.6 mag\,arcsec$^{-2}$, which corresponds to an IGrL fraction of $\lesssim 1\%$ of the total group light. This stringent upper limit suggests that IGrL does not contribute significantly to the Extragalactic Background Light at short wavelengths. Furthermore, the lack of UV IGrL observed in these stacks suggests that the atomic gas observed in the intragroup medium (IGrM) is likely not dense enough to trigger star formation on large scales. Future studies may detect IGrL by creating similar stacks at longer wavelengths or by pre-selecting groups which are older and/or more dynamically evolved similar to past IGrL observations of compact groups and loose groups with signs of gravitational interactions.

\end{abstract}



\section{Introduction} \label{sec:intro}

The hierarchical structure of galaxy formation predicts that large structures in the Universe form through the merging of smaller halos which began as small overdensities \citep{White1978,White1991,Springel2005,DeLucia2006}. As a result, the majority of galaxies today are observed to be located in group environments which reside in dark matter halos with masses between 10$^{12}$--10$^{14}$\,\msun \citep{Tully1987,Karachentsev2004}. Groups have therefore been of particular interest in hopes of understanding how group environments impact galaxy evolution. In order to fully characterize group environments, the gas gravitationally bound to the dark matter halo, termed the intragroup medium (IGrM), must be observed and understood. 

For higher mass halos, the IGrM has been observed through X-ray emission \citep{Mulchaey1996Xray,helsdon_ponman00,Mulchaey2000}; however, the majority of groups do not have gravitational potentials sufficient for X-ray emission from gas at their virial temperature. Quasar (QSO) absorption lines provide an alternative means of observing the IGrM. Despite being limited by chance alignment, QSO absorption lines allow for diffuse gas to be observed independent of redshift and only limited by spectral signal to noise and gas column density. \citet{Mulchaey1996} predicted that gas at the virial temperature of typical groups could be traced by broad, shallow \ovi\ absorption with the absence of cooler transitions such as \cii, \siii, and \nv. Early studies by \citet{Tripp2000}, \citet{TrippSavage2000} and \citet{Stocke2006} observed \ovi\ in IGrM sightlines, but the data were insufficient to correlate \ovi\ with the IGrM. Additional IGrM surveys by \citet{Stocke2019} and \citet{McCabe2021} concluded that the dominant, volume filling component of the IGrM should be traced by higher ionization transitions and that \ovi\ can be evidence of multiphase gas within the group halos. While QSO absorption lines have provided the easiest means of observing the IGrM, they are severely limited by both the number of QSO sightlines aligned behind groups and the absence of any transverse spatial information which direct imaging provides.

Another component of the IGrM, intragroup light (IGrL), has been observed in compact groups of galaxies, where the IGrM is expected to be heavily influenced by galaxy interactions and increased dynamical friction resulting from the lower velocity dispersion of the group members \citep[and references therein]{Hickson1977, Barnes1985,Purcell2007}. Combinations of tidal processes and interactions are thought to remove stars from galaxies and strand them within the larger group halo. Since these stars remain bound to the group, IGrL begins to build up slowly \citep{Mihos2004,Murante2004,Purcell2007}. As a result, Hickson Compact Groups \citep[HCG;][]{Hickson1982,Hickson1992} provided an ideal environment to observe IGrL due to the combination of high density and low velocity dispersions. Early studies by \citet{Nishiura2000} and \citet{White2003} found evidence of diffuse IGrL in HCGs 79 and 90, which corresponds to 13\% and 45\% of the total group light, respectively. Similarly, \citet{DaRocha2005} acquired deep B- and R- band images of HCGs 79, 88 and 95 and found a significantly higher fraction of the total group light in the IGrL of HCG 79 of 33\% in the R-band and 46\% in the B-band. However, for HCG 95, the IGrL was much less prominent as it was only 11\% and 12\% in the B and R bands respectively. No IGrL was detected in HCG 88 down to surface brightness levels of 29.1~mag\,arcsec$^{-2}$ in the B-band and 27.9~mag\,arcsec$^{-2}$ in the R-band. As a result, \citet{DaRocha2005} concluded that the IGrL becomes more prominent as the groups become more dynamically evolved. 

\citet{DaRocha2008} extended this analysis to three more Hickson Compact Groups and found similar results. For HCGs 15, 35 and 51, the fraction of IGrL ranged between 15-31\% in the B-band and 11-28\% in the R-band and the same correlation between IGrL fraction and indications of dynamical evolution appeared to exist. The B-R colors of the IGrL in HCGs 35 and 51 were determined to be bluer than the colors of the member galaxies, which may indicate that star forming regions, either in-situ or from tidal stripping, have a strong contribution to the IGrL. Other studies of HCGs by \citet{Aguerri2006} and \citet{Poliakov2021} found similar fractions of IGrL down to surface brightness levels of $\sim$28~mag\,arcsec$^{-2}$ in the $r$-band and an upper limit of 4.7\% of the group light for HCG 44 to 30.04~mag\,arcsec$^{-2}$ in the B-band. In each of these studies, the process in which interactions and tidal stripping lead to a build up of IGrL has been supported by observed correlations between the amount of IGrL and the fraction of early type galaxies in compact groups \citep{Aguerri2006,DaRocha2008,Poliakov2021}.

Expanding the search for IGrL to looser groups, \citet{Watkins2014} searched for IGrL in the M96/Leo I group. Using a combination of B- and V-band images from the Burrell Schmidt Telescope on Kitt Peak, no diffuse IGrL was observed down to surface brightness levels of $\mu_{B}=30.1$\,mag\,arcsec$^{-2}$. As a result, the authors suggest that frequent interactions are necessary to produce IGrL and that the M96 group is either not in an evolved enough state, too low in mass, or lacking a sufficient density of galaxies required for frequent/strong interactions. 

\citet{Spavone2018} and \citet{Raj2020} used the Fornax Deep Survey to study the IGrL of the loose groups NGC 5018 and Fornax A, respectively. The NGC 5018 group was found to have significant amounts of IGrL that constitute 41\% of the total group light in the $g$-band. The $g-r$ color of the detected IGrL is consistent with the member galaxies, suggesting that tidal interactions are primarily responsible for its formation. \citet{Raj2020} observed significantly smaller fractions of IGrL as only 16\% of the total group's $g$-band light was in the form of IGrL. Similarly to \citet{Spavone2018} and studies of compact groups by \citet{Aguerri2006,DaRocha2008} and \citet{Poliakov2021}, the authors believe that the IGrL observed was a result of tidal interactions from the disruption of dwarf galaxies in the group.

\citet{Cattapan2019} studied the dynamically young, un-virialized Dorado group and observed IGrL out to surface brightness limits of 30.11 mag\,arcsec$^{-2}$ in the $g$-band and 28.87 mag\,arcsec$^{-2}$ in the $r$-band. It was determined that tidal interactions were the cause for the build up of IGrL in this group. At higher redshifts, \citet{Martinez2023} detected IGrL to limiting surface brightnesses of 30.76 mag\,arcsec$^{-2}$ and 29.82 mag\,arcsec$^{-2}$ in the $g$ and $r$-bands, respectively, which corresponds to IGrL fractions between 2 and 36\%. This group was found to be dominated by early-type galaxies, which indicates that the group is dynamically evolved. 

To date, there has not been any statistical survey of IGrL across more typical groups across a single field. While compact groups represent the optimal case for IGrL detection and analysis, they do not represent the majority of groups in the universe due to the inherent population, isolation, and density requirements \citep{Hickson1982}. Therefore, a combination of deep imaging combined with a robust galaxy group catalog is necessary for a statistical study of IGrL without bias towards bright, dense, and rich groups. Suitable deep, ground based imaging was secured in support of UVCANDELS (PI H. Teplitz), which aimed to provide UV coverage of the Cosmic Assembly Near-Infrared Deep Extragalactic Legacy Survey \citep[CANDELS;][]{Grogin2011,Koekemoer2011} fields: Great Observatories Origins Deep Survey \citep[GOODS;][]{Giavalisco2004} North and South, the Extended Groth Strip \citep[EGS;][]{Davis2007}, and the Cosmic Evolution Survey \citep[COSMOS;][]{Koekemoer2007, Scoville2007}.

As part of the GOODS-N UVCANDELS observations, \citet{Ashcraft2018} used the capabilities of the Large Binocular Camera (LBC) on the Large Binocular Telescope (LBT) to complement the \textit{Hubble Space Telescope} (HST) parallel WFC3/UVIS F275W and ACS/WFC F435W observations with ground based \Ub ($\lambda_c\,\simeq359$ nm; $\Delta\lambda\simeq54$ nm) imaging. \citet{Ashcraft2018} pioneered the seeing sorted stacking method, which was used by \citet{Otteson2021}, \citet{Redshaw2022} and \citet{Ashcraft2023} to stack individual exposures (starting with the best seeing) incrementally to create ``optimal depth'' and ``optimal resolution'' mosaics. The optimal resolution and optimal depth stack FWHM cutoffs were dependent upon the seeing distributions for each individual exposure, but only the best $\sim 10\%$ of the exposures were used for the optimal resolution mosaic and only the worst $\sim 5-10\%$ were excluded from the optimal depth mosaic. This method allows the brightest galaxies to be studied through the optimal resolution mosaics, where large features appear to be resolved. Additionally, faint regions with low surface brightness (i.e. faint outskirts, tidal tails, plumes, clumps, etc.) are more easily studied in the slightly deeper, optimal depth mosaics, where an increased number of photons outweighs the need for higher resolution. Furthermore, since adaptive optics are not possible in the $U$-band, this method is necessary to mitigate the variable atmospheric effects that are present, even at an excellent observing site \citep{Taylor2004}.

\begin{figure*}
\centering
\plotone{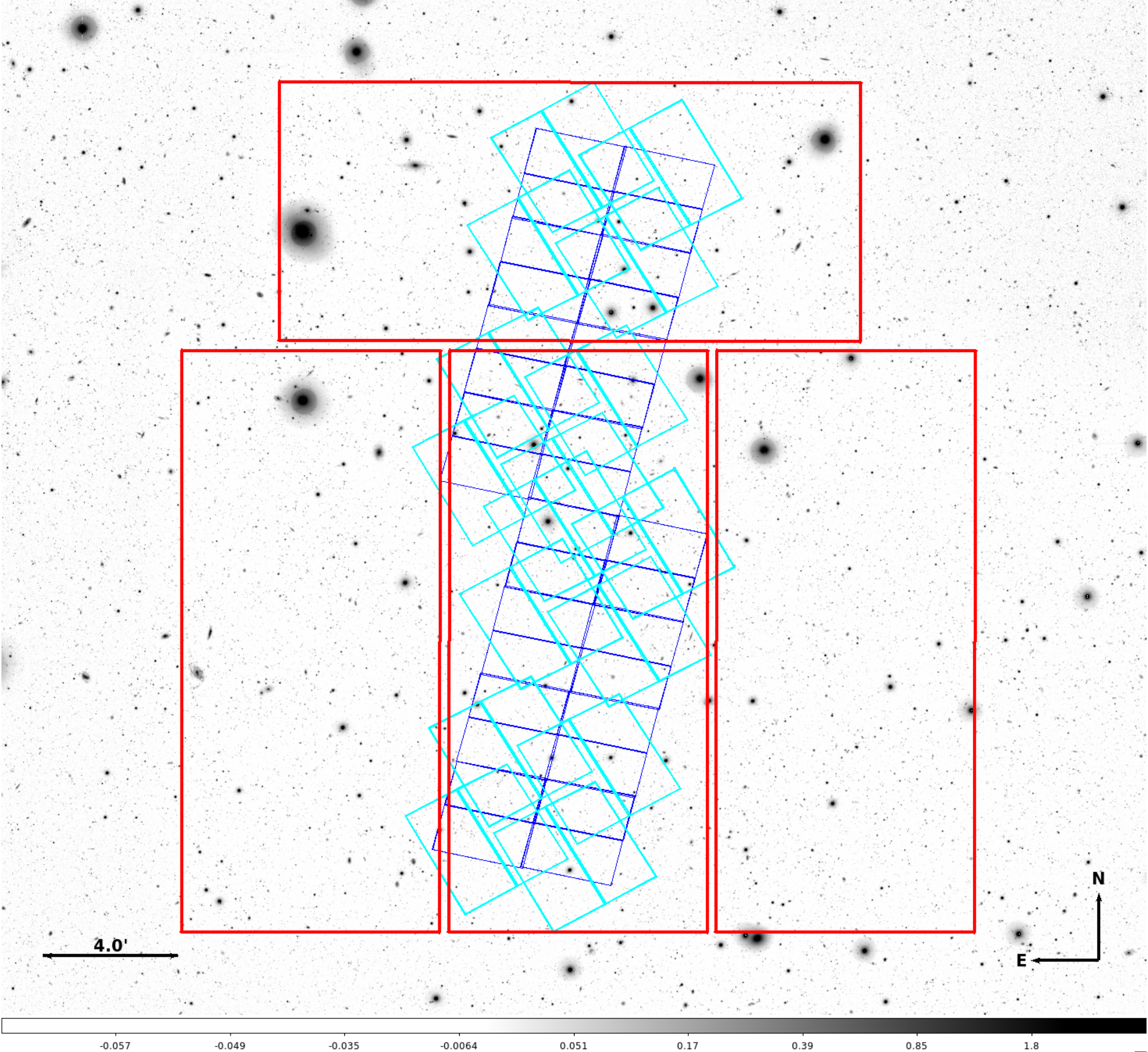}
\caption{A subsection of the full LBT/LBC $U$-band optimal depth COSMOS mosaic showing the UVCANDELS sub-region. The cyan and blue outlines represent the \textit{HST}/ACS and \textit{HST}/WFC3 footprints, respectively. These regions fit within a single LBC pointing, which is outlined in red. }
\label{fig:cosmos}
\end{figure*}

In order to search for IGrL, we will use the LBT/COSMOS \Ub optimal depth mosaic in conjunction with the zCOSMOS 20k galaxy group catalog \citep{Knobel2012} to identify and stack group backgrounds. While COSMOS is a large, $\sim$2 square degree field centered at R.A.$=$10:00:28.6 and Dec$=$+02:12:21, we limit our analysis to the central area of the field, which contains the UVCANDELS footprint (see Figure \ref{fig:cosmos}).

This paper is organized as follows: in \S \ref{sec:obs}, we describe the LBT observations, and in \S \ref{sec:mosaics} we discuss the stacking procedure along with the creation of the object catalogs. \S \ref{sec:igl} describes the search for diffuse, intragroup light in the UV through galaxy group stacks. Lastly, in \S \ref{sec:discussion}, we summarize our results and provide future outlook towards the search for IGrL. Unless stated otherwise, all magnitudes listed in this paper are in the AB system \citep{Oke1983} and Planck 2018 cosmology is adopted \citep{Planck2020}.

\section{Observations} \label{sec:obs}
We used the Large Binocular Cameras \citep[LBCs;][]{Giallongo2008} to obtain 532 observations of the COSMOS field. The twin, wide-field instruments at the prime foci of the LBT each have a $\sim23\farcm6\times25\farcm3$ field of view (FoV) and are able to simultaneously observe the same target with red and blue optimized CCDs. The LBC-red camera is optimized for the $V$--$Y$ bands ($500-1000$ nm), while the LBC-blue camera is optimized for the UV--$R$ bands ($350-650$ nm) which are the focus of this paper. The two cameras contain four $4\,\rm{K}\,\times\,2\,\rm{K}$, E2V 42-90 CCDs, which are characterized by: a gain of $\sim$\,1.75\,$e^{-}$/ADU, read-noise of $\sim\,$9 ADU, and a plate scale of $\sim0\farcs2254$\,pix$^{-1}$. For all LBC observations discussed in this paper, the SDT $U_{\rm spec}$ filter was used. The bandpass of this filter when combined with the detector quantum efficiency and the telescope/instrument optics is characterized by a central wavelength of $\lambda_c = 3590$ $\angstrom$, a band width of 540 $\angstrom$, and a peak throughput of $\sim$38\% \citep[Figure 2 of ][]{Giallongo2008}.

The majority of these LBC observations were obtained between 2007 and 2014, while an additional set of observations were carried out in December 2019 and January 2020 in support of UVCANDELS. In total, 532 individual exposures were obtained of the entire COSMOS field, with the newest 94 focusing on the UVCANDELS region. Each observation was uniquely dithered, so that the detector gaps were adequately covered and cosmic rays could be robustly rejected. All images were processed by the Italian LBT partners, involving bias subtraction, flat-fielding, and astrometric calibration onto GAIA/DR3, through the LBC reduction pipeline as detailed in \citet{Giallongo2008}.

\section{Mosaic and Catalog Creation} \label{sec:mosaics}
\subsection{Floating Zero Point Correction}
\label{sec:transp}

The GOODS-N optimal seeing and optimal depth mosaics created in \citet{Ashcraft2018} were found to have slight differences in their photometric zero points. While these differences were small, $\sim$0.2~mag, an additional step was added to the seeing sorted stacking method in order to correct for this offset, which was speculated to be a result of varying transparency and sky brightness atop Mt. Graham \citep{Taylor2004,Ashcraft2018}. In order to address this slight offset, we investigated the relative transparency of each of the 532 exposures and corrected for any offset in the zero points following the procedures described in \citet{Otteson2021}, \citet{Redshaw2022} and \citet{Ashcraft2023}. For each individual exposure, a subset of $\sim150$ unsaturated stars was selected from a Sloan Digital Sky Survey (SDSS) Data Release 16 \citep{Blanton2017,Ahumada2020} based catalog of stars with \Ub magnitudes, $18 \le u \le 22~\rm{mag}$. The flux of these unsaturated stars was calculated and compared to the flux from the same sources in a single LBT exposure. The 3$\sigma$ clipped median value of the flux ratios for all $\sim$150 stars was adopted as the relative atmospheric transparency value for a single exposure. 

\begin{figure}
\centering
\plotone{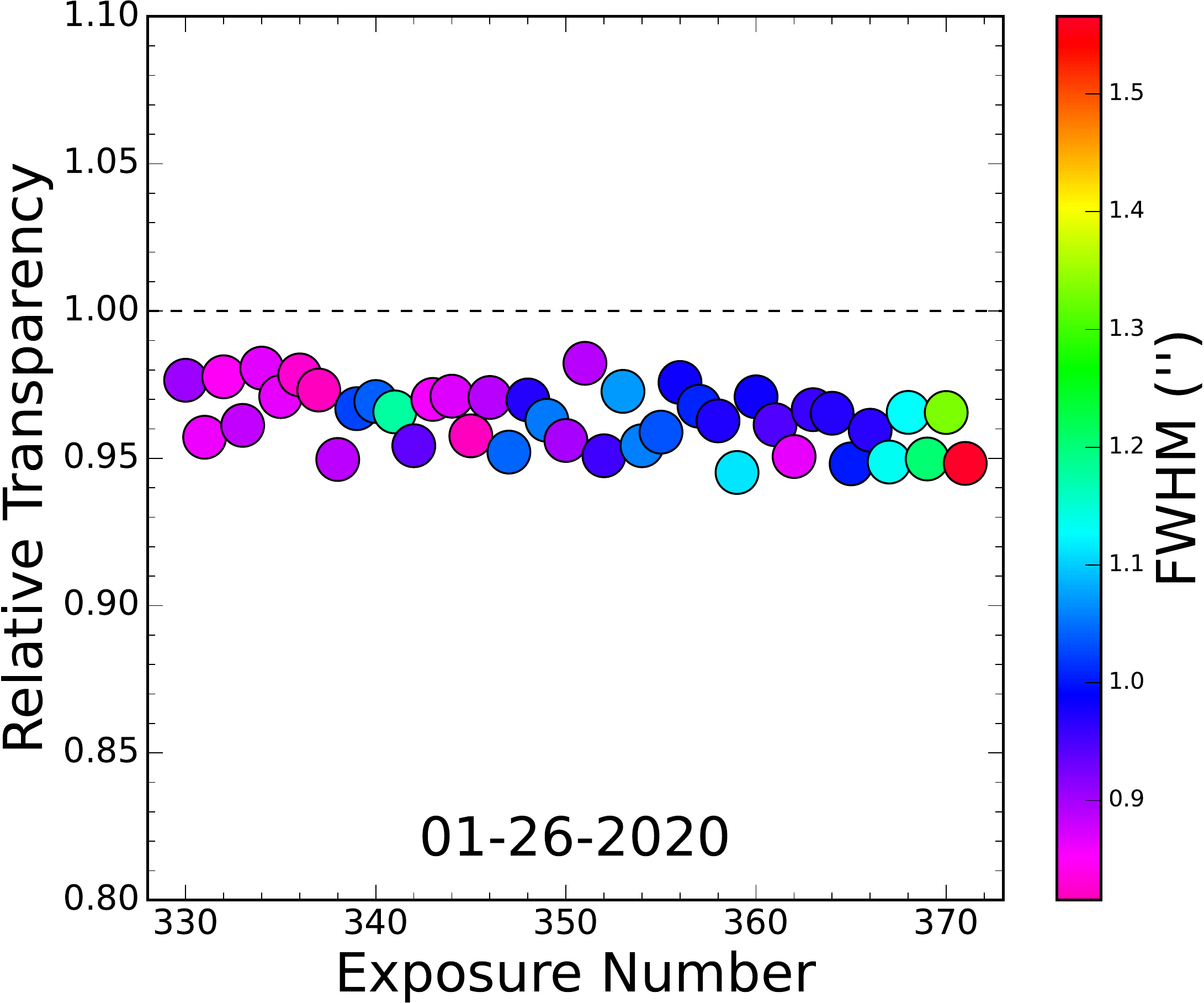}
\caption{Relative transparency distributions for one night of data in January 2020. The color bar represents the median seeing for each observation and the black, dashed line represents an atmospheric transparency equal to that of SDSS. }
\label{fig:transp}
\end{figure}

Figure \ref{fig:transp} shows the relative transparency values for each exposure taken on January 26, 2020. This distribution shows a median atmospheric transparency of $\sim$96\%. While this night's atmospheric transparency differed from the SDSS catalog by only a few percent, this difference in atmospheric throughput is compounded when stacking hundreds of individual exposures. Therefore, a correction was applied to each exposure so that the atmospheric transparency, and thus the photometric zero point, was equal to that of SDSS. After this correction, the transparency for each exposure will be at unity in Figure \ref{fig:transp}. This process was repeated for each of the 532 exposures in our dataset.

\subsection{Seeing Sorted Stacks}
\label{sec:seeing}

\begin{figure}
\centering
\plotone{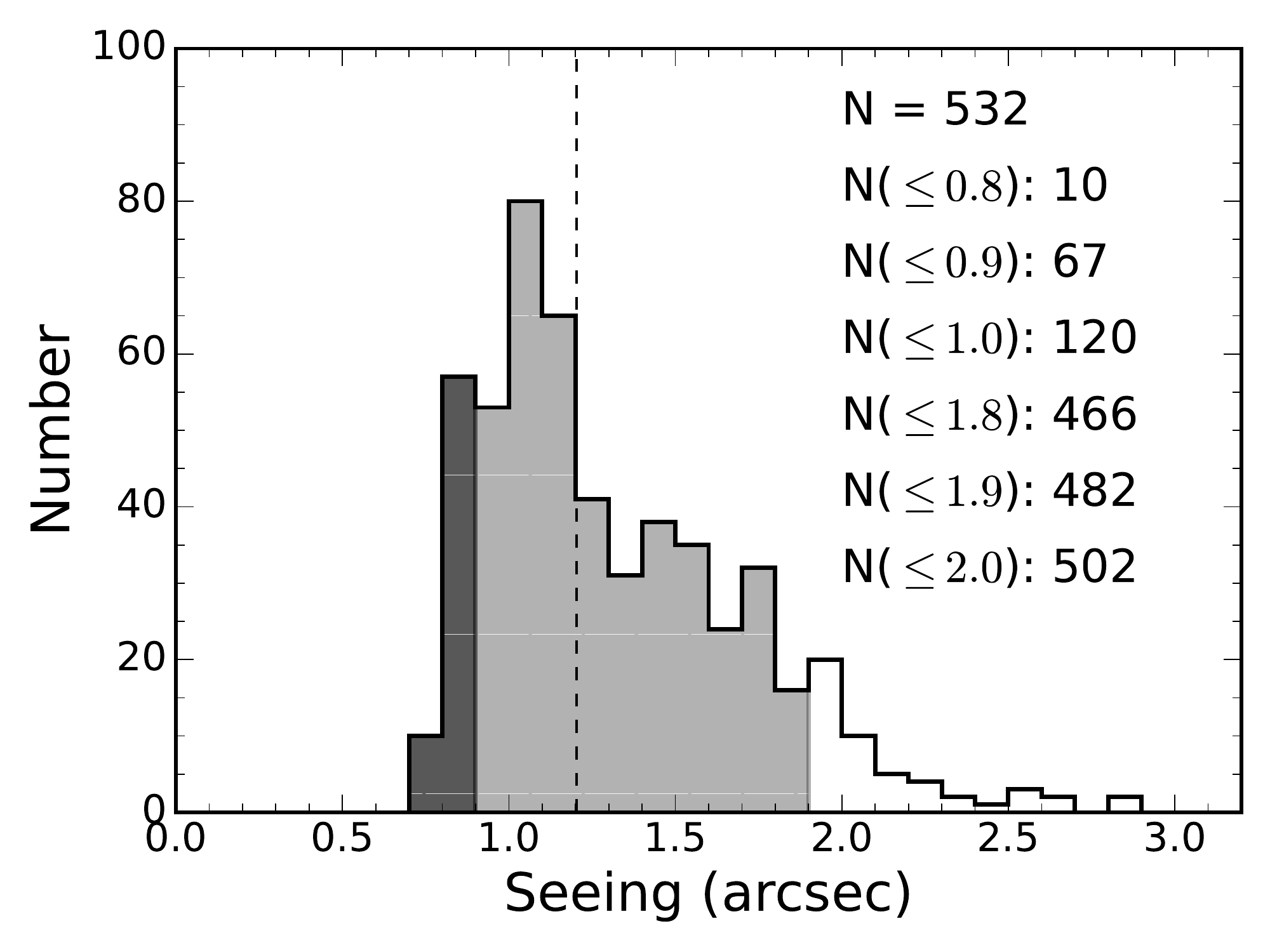}
\caption{Histogram of the FWHM measurements for each of the 532 usable, individual exposures in the COSMOS field. The vertical, dotted line represents the median seeing ($1\farcs2$ FWHM) of the dataset. The dark and light shaded regions represent the subset of exposures used for the optimal resolution and optimal depth mosaics, respectively. }
\label{fig:seeing_hist}
\end{figure}

For all 532 exposures, the individual LBC tiles were combined using \swarp \citep{Bertin2002,Bertin2010}, so that \sextractor \citep{Bertin1996} could be used to measure the Gaussian FWHM of $\sim100$ unsaturated stars in the image. The median of the FWHM distribution for the subset of unsaturated stars was used as the seeing of the entire exposure. The seeing was compared to the tabulated value from the data reduction pipeline, and since they were in agreement, the pipeline FWHM was used for each image. Once the seeing of each image was determined, the 532 observations were sorted as a function of seeing as shown in Figure \ref{fig:seeing_hist}. The median seeing value was determined to be $\sim1\farcs2$ FWHM, which is slightly higher than the value of $\sim1\farcs1$ FWHM, reported by Ashcraft et al. (2018) for the GOODS-N field. However, this still warranted creating separate optimal depth and optimal resolution stacks. 

\begin{figure*}
\centering
\includegraphics[scale=0.63]{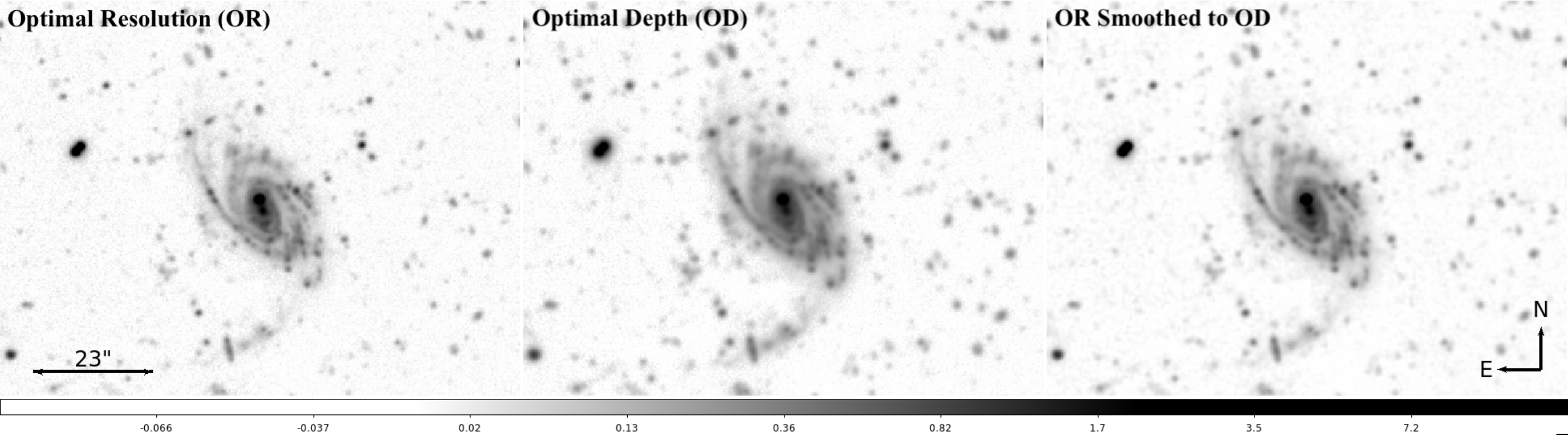}
\caption{Comparison of a $\sim 100'' \times 75''$ subsection of the optimal resolution (left) and optimal depth (middle) mosaics for the LBT/LBC data. The best seeing stack is composed of 67 individual exposures with FWHM$\le0\farcs9$, while the optimal depth mosaic is created from a stack of 482 image with FWHM $\le1\farcs9$. The right panel shows the optimal resolution mosaic smoothed to the seeing of the optimal depth mosaic. Despite being smoothed, this cutout does not capture the low surface brightness flux that is observed with the optimal depth mosaic, which reinforces the need for both optimal resolution and depth stacks.}
\label{fig:mosaics}
\end{figure*}

The COSMOS data were combined into two stacks with seeing less than $0\farcs9$ FWHM (optimal resolution) and seeing less than $1\farcs9$ FWHM (optimal depth). \swarp was used to perform the image stacking using 5$\sigma$ clipping and LANCZOS3 resampling on datasets of 67 and 482 images for the optimal resolution and depth mosaics respectively. An example of the two stacks is shown in Figure \ref{fig:mosaics}, where the left panel is the optimal resolution and the middle panel is the optimal depth mosaic. In the best depth stack, the low surface brightness tidal tail of the central galaxy is more prominent, while the individual spiral arms and star forming regions can be best discerned in the optimal resolution stack. The necessity of both stacks is highlighted through smoothing the optimal resolution stack to the seeing of the optimal depth stack, where the faint tidal tail is not detected to the level of the optimal depth mosaic. \\

\subsection{\Ub Catalogs}
\label{sec:catalogs}

Once the optimal depth and optimal resolution mosaics were created, catalogs were constructed for all of the sources in the mosaics. The catalogs were created using \sextractor and configuration parameters similar to those used in \citep{Ashcraft2018}. The sky background was determined through the use of a 6$\times$6 pixel median filter and a 256$\times$256 pixel and 128$\times$128 pixel mesh for the optimal resolution and depth mosaics, respectively. The sources were detected using a Gaussian filter and a 5$\times$5\,pixel convolution kernel with a FWHM of 3\,pixels. 

The photometric zero point was verified to be 26.50$\pm$0.11\,mag and 26.45$\pm$0.11\,mag for the optimal resolution and optimal depth mosaics, respectively, by comparing with the photometry of stars between $U_{AB}=18$\,mag and $U_{AB}=22$\,mag within the SDSS catalog. The consistency of these zero points reflects the improvement that was made through the use of the atmospheric transparency corrections.

\begin{figure}[]
\centering
\plotone{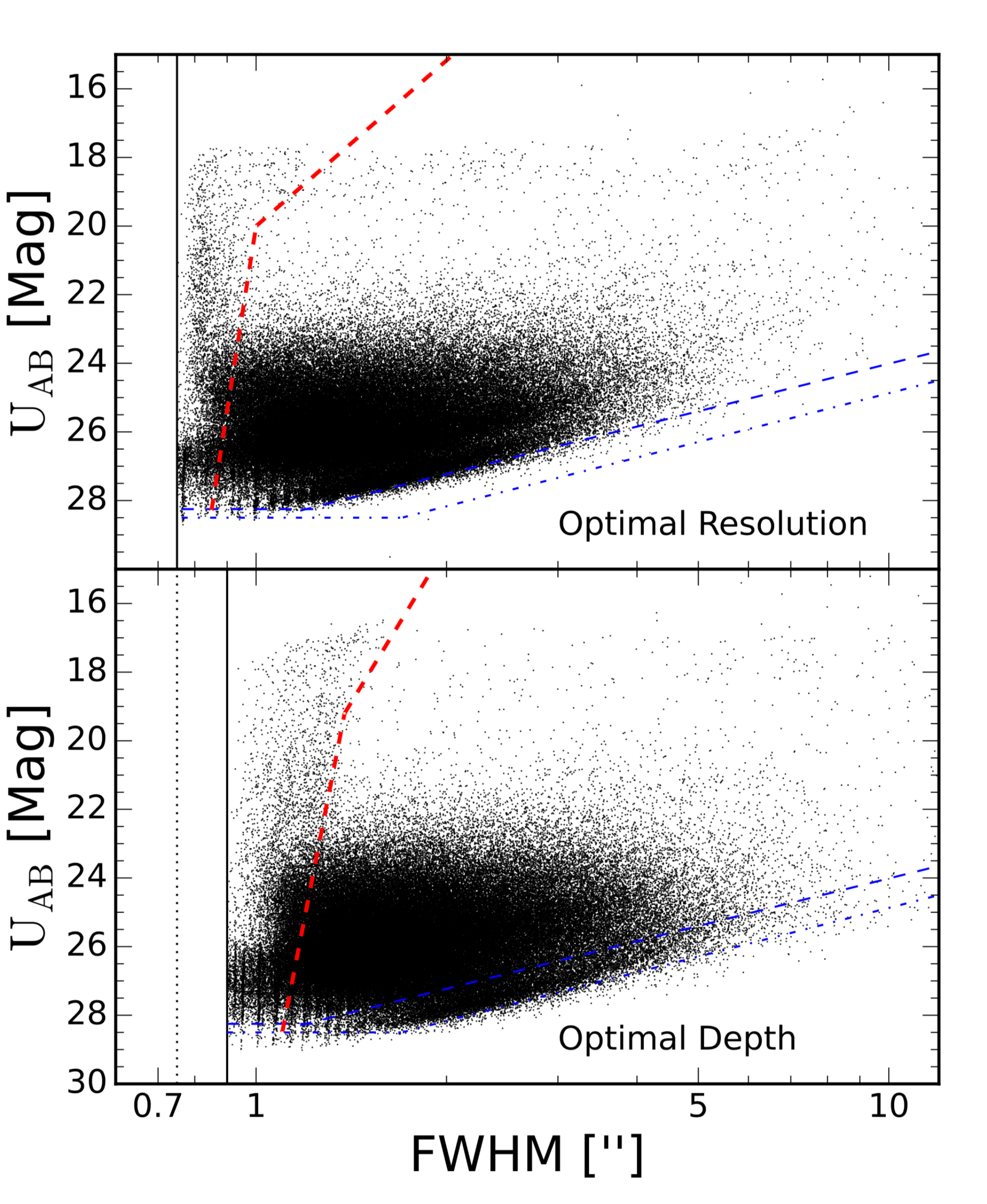}
\caption{Object FWHM as a function of U-band magnitude for the optimal resolution and optimal depth mosaics. The red, dashed lines show the star/galaxy separation and the solid, black lines illustrate the lower FWHM limits. The blue, dashed and dot-dashed lines represent the surface brightness limits for the optimal resolution and depth mosaics, respectively.}
\label{fig:elephant}
\end{figure}

In order to identify the stars in the source catalog, all sources were placed on a magnitude vs.\ FWHM diagram, shown in Figure \ref{fig:elephant}, where the non-saturated stars occupy a vertical strip at FWHM $\sim1\farcs2$, the $U$-band median seeing of the dataset. Saturated stars are also easily identified through the ``curled'' tip of the vertical strip of stars. To the right of $1\farcs4$ FWHM, one finds the galaxies identified in the optimal depth stack. Knowing how stars and galaxies populate different regions of the magnitude vs.\ FWHM-size diagram allows one to construct separate differential number counts for stars and galaxies.

Typically, the magnitude vs.\ seeing distribution for a single LBT/LBC pointing should show a much narrower vertical strip of unsaturated stars along with the same ``curled'' region for saturated stars. However, due to the large area of the COSMOS field on the sky, it takes four LBC pointings to cover the entire field. Prior to UVCANDELS, there was no high priority region within the field for deep HST observations (see Figure \ref{fig:cosmos}), and early LBT observations were spread over the entire two square degree field and on nights with different seeing conditions. 

\begin{figure*}[]
\centering
\plottwo{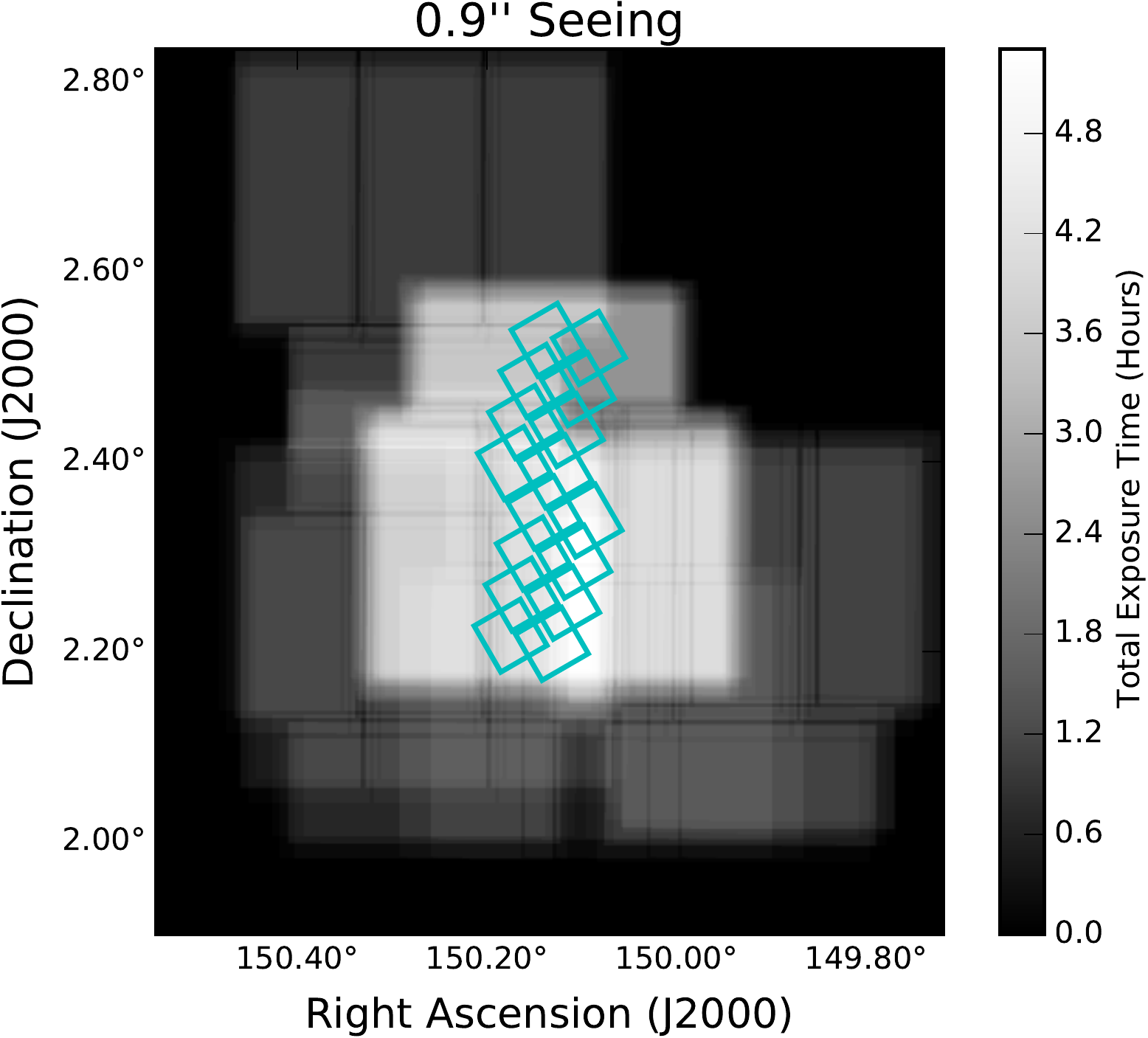}{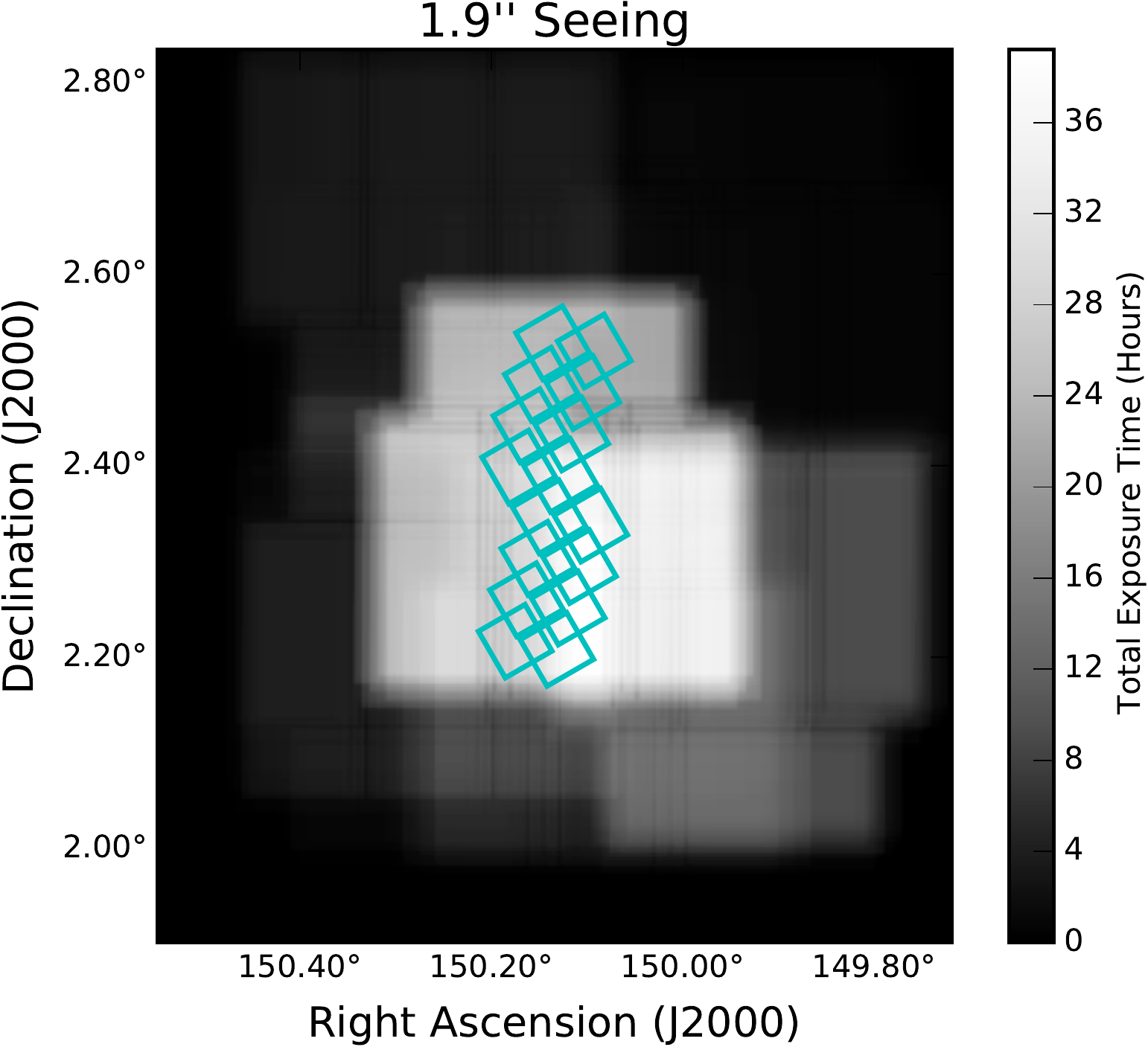}
\caption{LBC exposure maps showing the total integration time per area on the COSMOS field. The left panel shows the optimal resolution mosaic and the right shows the optimal depth mosaic. The UVCANDELS \textit{HST}/ACS footprint is represented by the cyan boxes. Note that the greyscale represents different exposure times in each panel.  }
\label{fig:coverage}
\end{figure*}

The non-uniform coverage causes the depth to vary as a function of position within the COSMOS field. Figure \ref{fig:coverage} shows the non-uniform coverage of the LBT COSMOS dataset for the optimal resolution (left) and the best depth (right) mosaics. We achieved a total exposure time of $\sim 5$ hours and $\sim 37$ hours for the optimal resolution and depth mosaics, respectively. For comparison, \citet{Ashcraft2018} achieved total exposure times of 3.2 hours and 30.7 hours for their $0\farcs8$ and $1\farcs8$ FWHM stacks in the GOODS-N field. 

As a result of the uneven coverage across the COSMOS field, we divided the two mosaics into three regions (shallow, medium, and deep) based on total exposure time per pixel. These regions were selected to maximize coverage of each region, while also maintaining connected areas for each region. Table \ref{tab:depths} details the three analysis regions for each mosaic and defines the exposure time cutoffs for each as well as the total area. These sub-catalogs allowed for number counts to be created for areas of the full field with approximate equal depth. 

\begin{deluxetable}{lCc}[ht!]
\tabletypesize{\scriptsize}
\tablecolumns{3}
\tablecaption{Zone definitions for non-uniform LBT coverage of COSMOS}
\label{tab:depths}
\tablehead{
\colhead{Regions}  & \colhead{Exposure Time (hr)} 	& \colhead{Area (sq. deg)} \\[-0.3cm]	
}
\startdata
\cutinhead{Optimal Resolution}
Shallow      &  1-2       &   0.206   \\
Medium       &  2-4.2     &   0.101   \\
Deep         &  >4.2      &   0.020   \\
\cutinhead{Optimal Depth}
Shallow      &  10-24     &   0.205   \\
Medium       &  24-32     &   0.050   \\
Deep         &  >32       &   0.010   \\
\enddata

\end{deluxetable}
\normalsize

\begin{figure*}
\centering
\plottwo{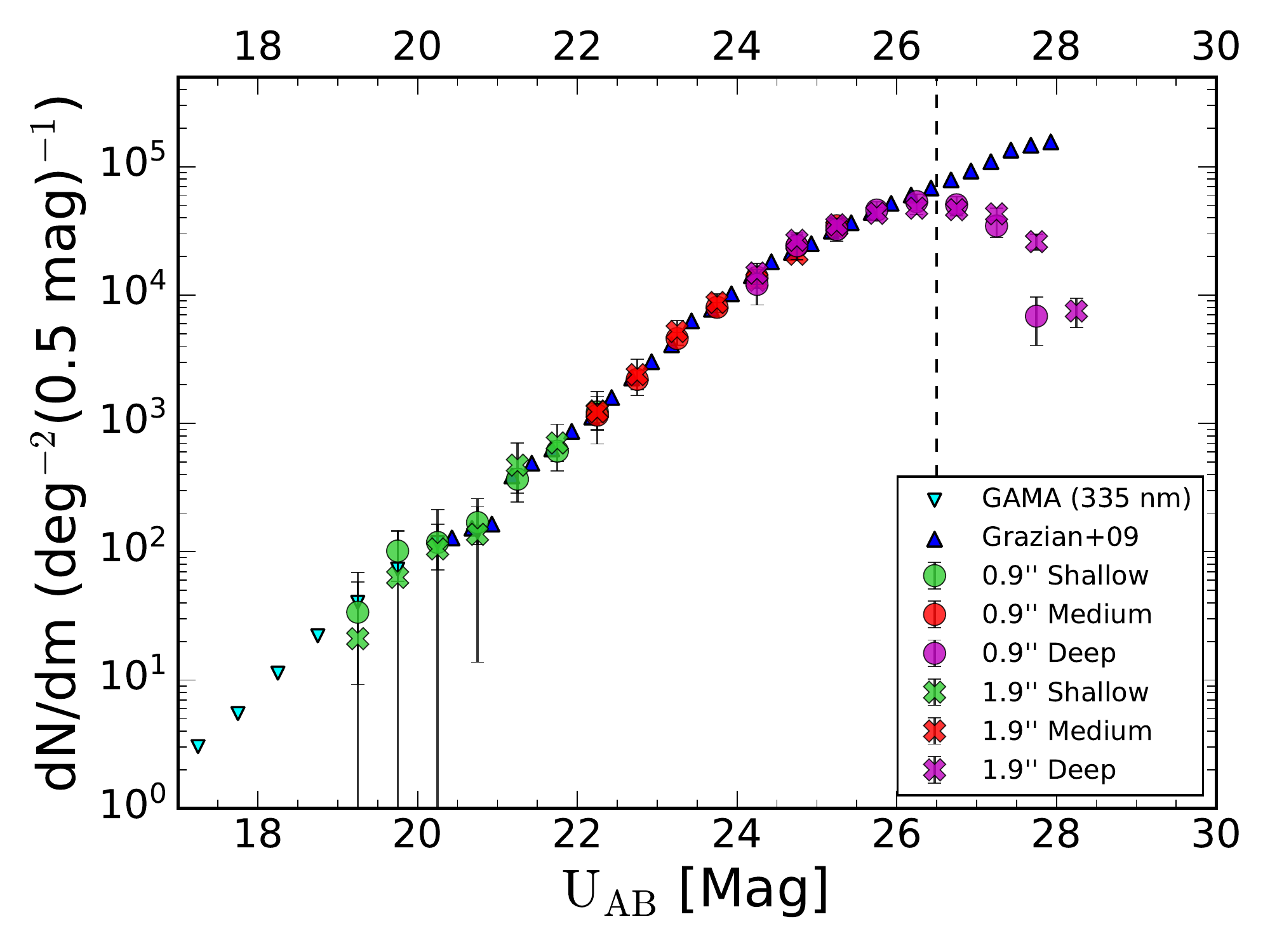}{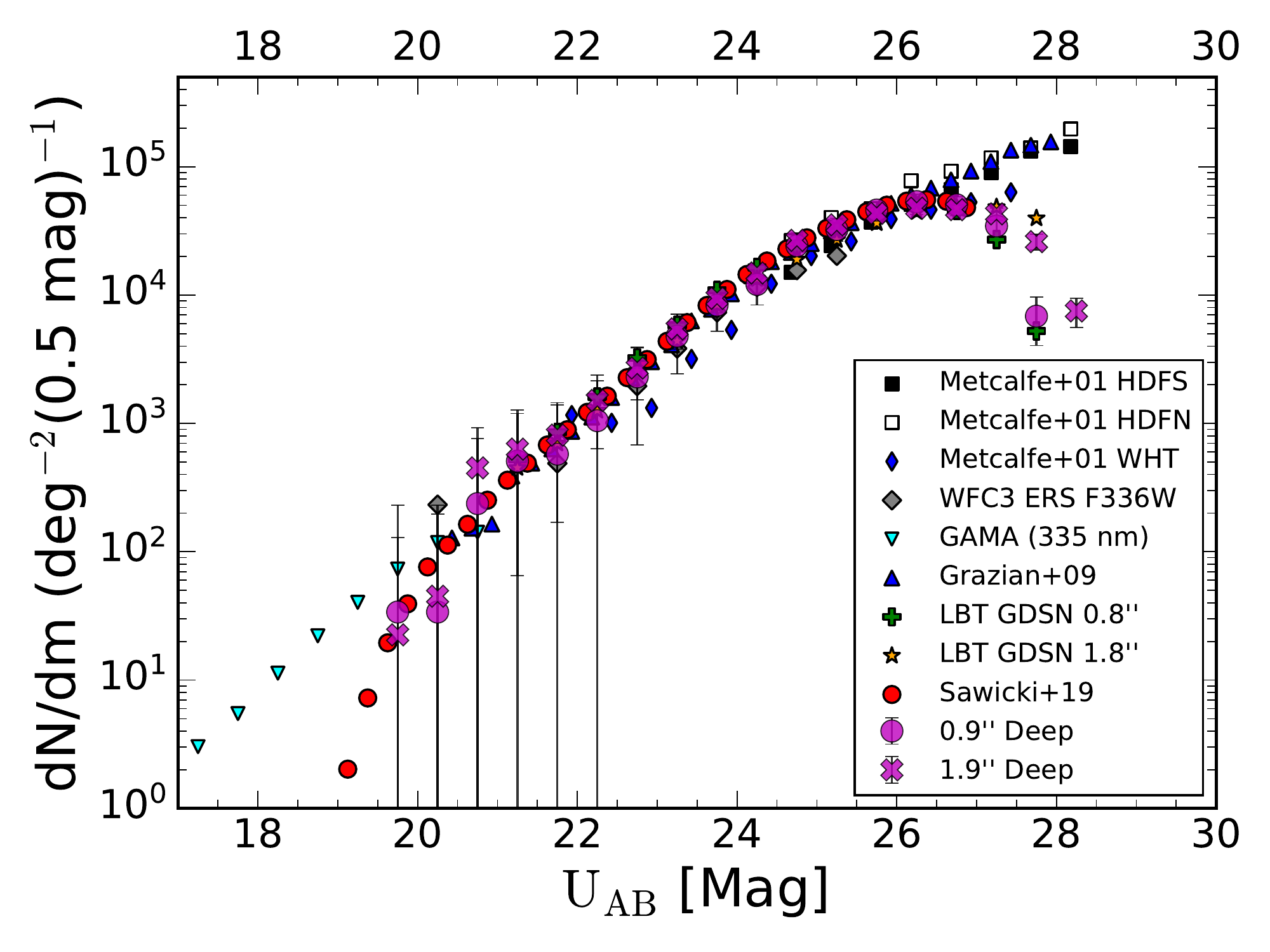}
\caption{\emph{Left: }Differential galaxy counts for the optimal resolution and optimal depth mosaics. The green, red, and magenta colors represent the shallow, medium, and deep regions, respectively. The vertical, dashed line represents the completeness limit of $\sim$26.5\,mag as the COSMOS counts begin to significantly deviate from the power law determined by the completeness corrected data from \citet{Grazian2009}.
\emph{Right: }The optimal depth and resolution galaxy counts in COSMOS compared to the counts in GOODS-N from \citet{Ashcraft2018} as well as various studies in other fields by \citet{Metcalfe2001}, \citet{Driver2009}, \citet{Grazian2009}, \citet{Windhorst2011}, and \citet{Sawicki2019}. The bright end of the COSMOS number counts deviate from the larger GAMA fields due to the selection against bright stars/galaxies in the definition of the COSMOS field \citep{Koekemoer2007,Scoville2007}. } 
\label{fig:counts}
\end{figure*}

Differential galaxy counts were created after separating the stars and galaxies following the prescriptions of \citet{Windhorst2011}. These galaxy counts, shown in Figure \ref{fig:counts}, show that the optimal resolution and optimal depth stacks reach U-band magnitudes of $U\sim$26 and $U\sim$26.5\,mag respectively. The depths reached in these mosaics are comparable to \citet{Ashcraft2018} (26\,mag across GOODS-N) as well as the CFHT Large Area \Ub\ Survey \citep[CLAUDS;][]{Sawicki2019}, which reached a depth of 26.3 mag over an area of $\sim$4 square degrees in the COSMOS field.

\section{Searching for Intragroup U-band Light} \label{sec:igl}
In order to search for diffuse light from the IGrM, we used the optimal depth mosaic in conjunction with the zCOSMOS 20k galaxy group catalog \citep{Knobel2012}. As such light is expected to be too faint to detect in individual groups, we aim to search for IGrL in a composite of multiple groups using image stacking. Three redshift ranges were selected for group stacking, each with increasing redshift and sample size: $0.1 < z \le 0.2$ (N=17), $0.15 < z \le 0.25$ (N=27) and $0.25 < z \le 0.35$ (N=33). We use all groups that are fully contained within the LBC mosaic, excluding groups in regions of the mosaic where the exposure map has a strong gradient which resulted in a varying background. Furthermore, only groups with N$>$3 spectroscopically confirmed group members were included to minimize the probability that an included group is only a chance superposition and not a physical group. These redshift slices were specifically selected to increase the number of groups per stack while also retaining enough area per group to perform the rescaling and stacking processes. 

\begin{figure*}
\centering
\includegraphics[scale=0.43]{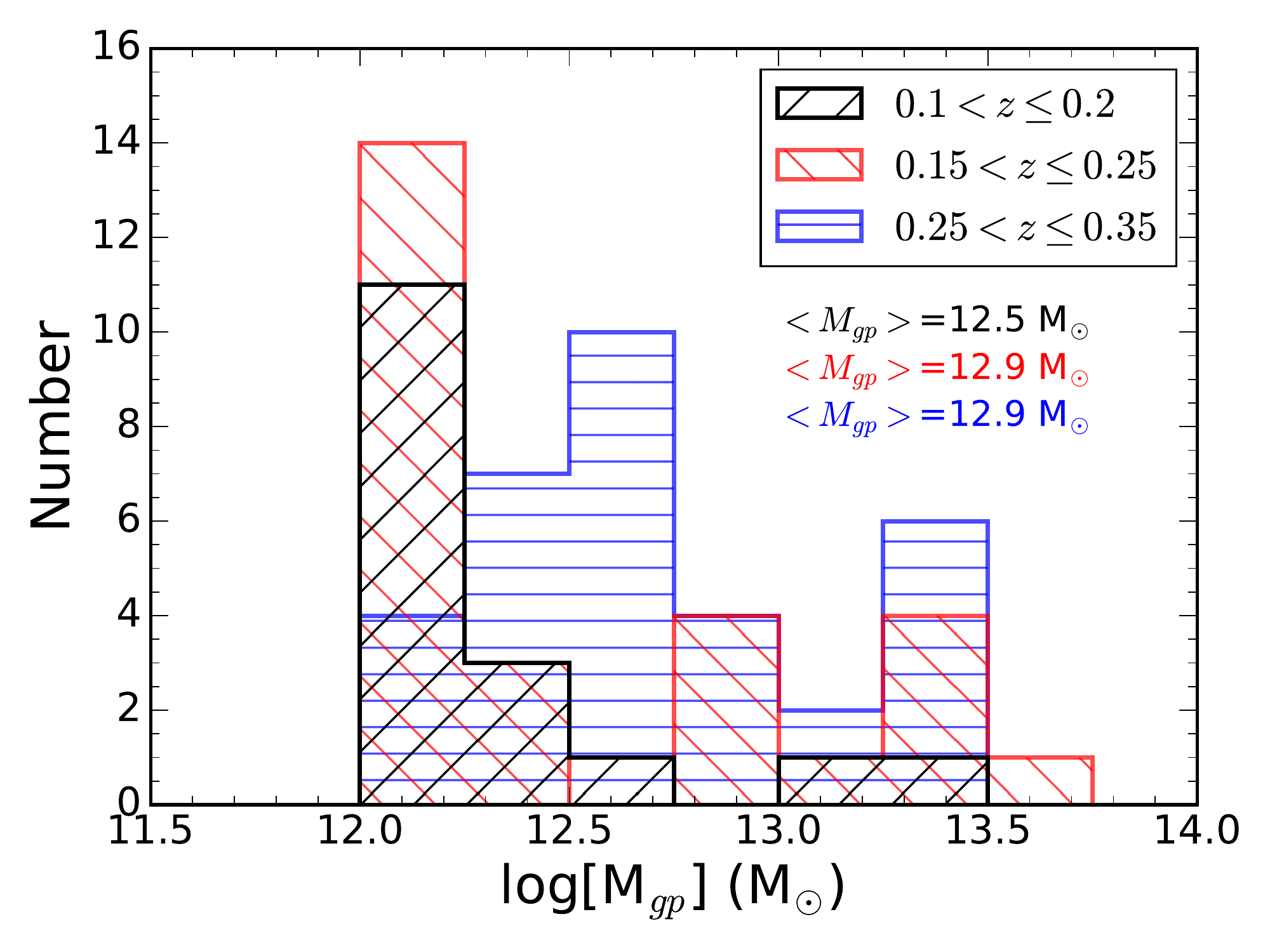}
\includegraphics[scale=0.43]{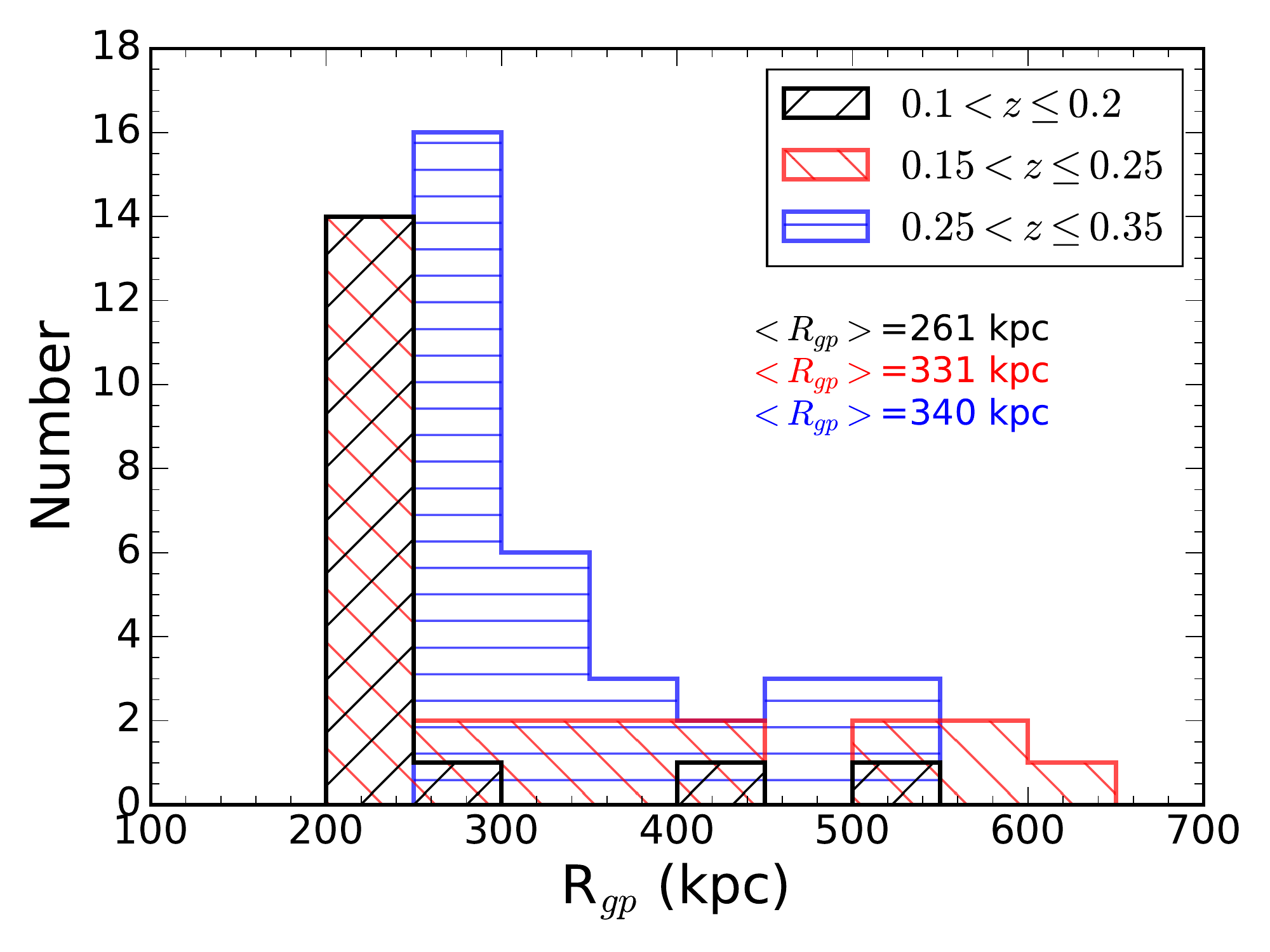}
\caption{Halo mass (left) and radii (right) distributions for each of the three subsets of groups in the sample. For reference, the average group radius for the lowest redshift bin has an average halo mass and radius which is in line with estimates for the Leo group \citep{Watkins2014} and the average group radii are 3--4 times as large as typical HCGs \citep{Poliakov2021}. This indicates that this sample probes loose groups, which are expected to be more common compared to more dense compact groups \citep[and references therein]{Coenda2012}. }
\label{fig:sample_hists}
\end{figure*}

As a sample, these groups do not have any bias towards those which are evolved or have experienced recent interactions. Upon visual inspection, these groups do not appear to be dominated by galaxy mergers or interactions. Figure \ref{fig:sample_hists} shows the halo mass (left) and group radii (right) distributions for each of the three redshift subsets of our group sample. On average, these groups are more representative of loose groups compared to the more commonly studied compact groups. The halo mass range of this sample of groups is typically between 0.5--1 dex from the loose groups used by \citet{Coenda2012} to study the different galaxy populations in loose and compact groups. Conversely, the group radii in this sample are typically 3--4 times as large as the radii for HCGs studied in \citep{Poliakov2021}.

Within each redshift range, four different stacks were constructed with two stacks corresponding to observed groups from the \citet{Knobel2012} catalog and the other two being random areas for a control sample. The random stack was created to mimic the observed group distributions as closely as possible, while existing at random positions within the optimal depth U-band mosaic. The random ``group'' redshifts and radii were randomly selected (without replacement) from the true group sample in order to preserve their inherent distributions. 

For each of the observed and random groups, each pixel area was rescaled by one of two methods: fractional or physical. The fractional stack was created by rescaling each group so that every group was the same number of pixels across. This allowed for each group's radius to be aligned such that corresponding fractional radii are stacked together. The physical stack was created by rescaling each group so that each group had the same physical size per pixel given their angular size distance, which allowed the same physical distance in each pixel to be stacked. In both cases, each group was rescaled using Lanczos resampling, such that resolution was always decreased and existing pixels were not split to increase resolution. Care was taken to ensure that flux was conserved between the original and rescaled group cutouts.

For each of the 12 stacks, the stacking process is described as follows: 
\begin{enumerate}  
    \item Group cutouts four group radii (4R$_{gp}$) in size on each side were created from the optimal depth mosaic. 
    \vspace{-0.2cm}
    \item Each group was rescaled via the fractional or physical methods to match the lowest resolution group in the sample. 
    \vspace{-0.2cm}
    \item Source Extractor was run on each cutout to generate segmentation maps, which were used to remove all sources to a level of 0.5$\sigma$ above the background.
    \vspace{-0.2cm}
    \item Due to slight gradients in the cutout backgrounds, the Source Extractor background map was fit with a plane and then subtracted from the group cutout. These gradients were measured over areas larger than the group sizes, which ensures that faint flux contributing to the IGrL was not subtracted from each group. 
    \vspace{-0.2cm}
    \item Each group was then tested for overlap with other groups in the sample. Any overlap was masked out (given zero weight) and not included in the stack. 
    \vspace{-0.2cm}
    \item Lastly, the remaining group background was coadded together for each group in the sample with equal weight. 
\end{enumerate}

\begin{figure}
\centering
\includegraphics[scale=0.4]{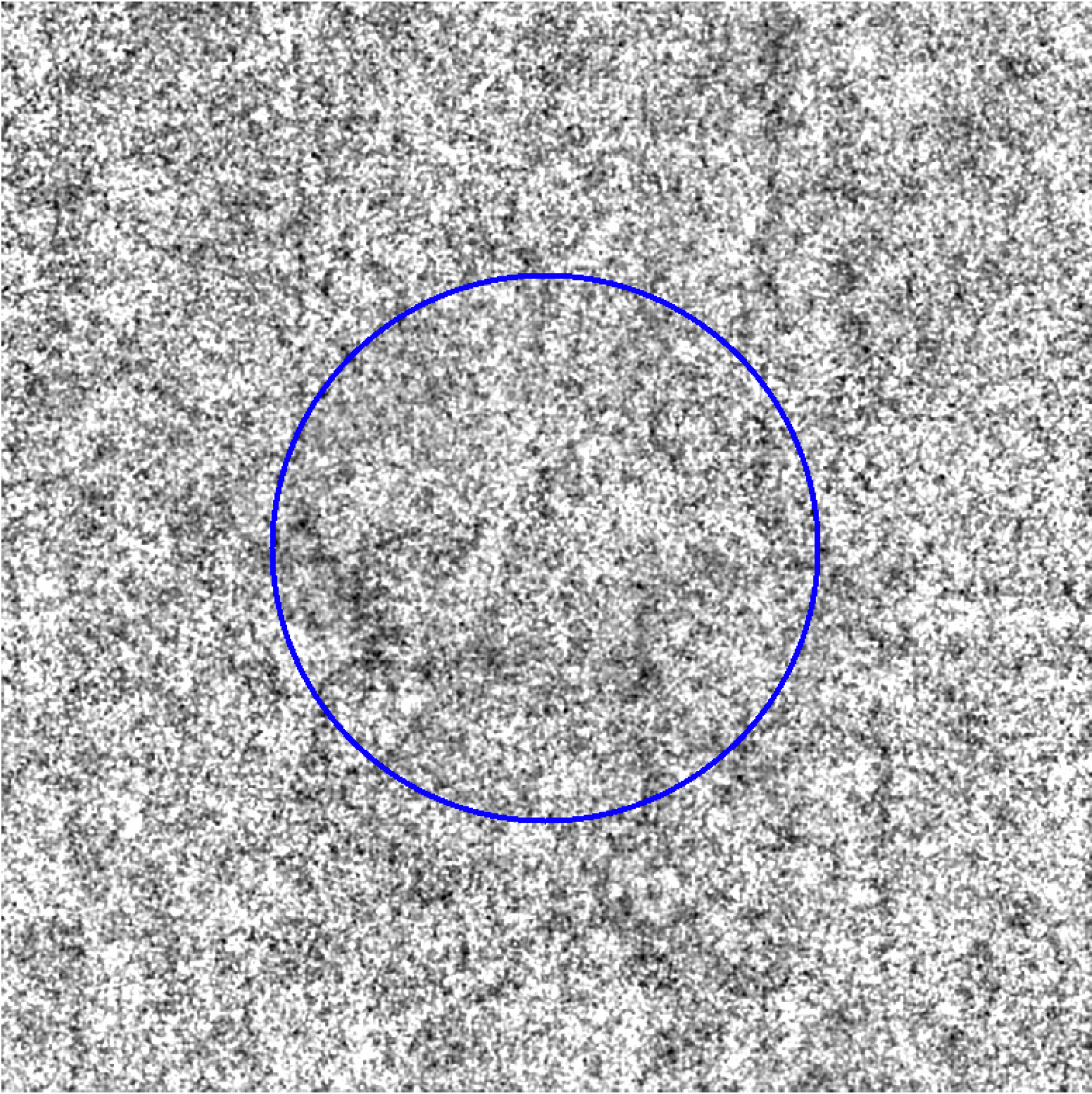}
\caption{Example of a group stack with 27 groups between redshifts $0.15 < z \le 0.25$. This stack was rescaled via the fractional method, where the blue circle represents the group radius for each group in the stack. There is no IGrL detected in this stack, which corresponds to a 3$\sigma$ upper limit of $\mathbf{29.17^{-0.33}_{+0.49}}$\,\rm{mag\,arcsec}$^{-2}$. }
\label{fig:stack}
\end{figure}

The group stacking results are summarized in Tables \ref{tab:fractional_stack} and \ref{tab:physical_stack} and an example stack is shown in Figure \ref{fig:stack} for the medium redshift stack. Across each stack, there was no IGrL detection as each appeared as random noise. For each of the observed group stacks, the rms of the stack was used to calculate a 3$\sigma$ IGrL upper limit to average surface brightness levels of $\sim 29.1-29.6$\,\rm{mag\,arcsec}$^{-2}$. A bootstrap analysis with replacement was used to determine the uncertainties in each IGrL upper limit. No significant difference was found between the observed and random group stacks, further reinforcing these IGrL non-detections in the observed U-band. There was also no significant difference found between stacking algorithms as the fractional and physical stacks resulted in similar surface brightness limits on the IGrL.

\begin{deluxetable*}{lccccccCc}[t]
\tabletypesize{\scriptsize}
\tablecolumns{7}
\tablecaption{Fractional Stack Measurements}
\label{tab:fractional_stack}
\tablehead{
\colhead{Stack Type}		&		\colhead{Redshift Range}	&	\colhead{$z_{\rm{med}}$}	&  \colhead{Number Groups}	&		\colhead{Mag} & \colhead{N$_{*}(>3M_\odot)$} &   \colhead{$\Sigma_{*}(>3M_\odot)$}	& \colhead{$\mu ^{AB}_{U}$} &   \colhead{Excess over Random} \\
\colhead{(1)}         			 & \colhead{(2)}       		 & \colhead{(3)}      			 &	\colhead{(4)}     & \colhead{(5)} & \colhead{(6)} & \colhead{(7)} & \colhead{(8)}     &   \colhead{(9)}
}
\startdata
Observed    &   $0.1 < z \le 0.2$       &   0.166   &   17  &   $>$ 32.39       & $<$3,100   &   $<$0.02   &   $>$29.14$_{+0.32}^{-0.24}$    &   -25.5\%   \\
Observed    &   $0.15 < z \le 0.25$     &   0.220   &   27  &   $>$ 32.42       & $<$5,500   &   $<$0.03   &   $>$29.17$_{+0.49}^{-0.33}$    &   18.0\%   \\
Observed    &   $0.25 < z \le 0.35$     &   0.332   &   33  &   $>$ 32.67       & $<$11,000  &   $<$0.03   &   $>$29.42$_{+0.28}^{-0.22}$    &   18.0\%   \\
\hline
Random      &   $0.1 < z \le 0.2$       &   0.166   &   17  &   $>$ 32.07       & $<$4,100   &   $<$0.02   &   $>$28.82$_{+0.26}^{-0.21}$    &   --  \\
Random      &   $0.15 < z \le 0.25$     &   0.220   &   27  &   $>$ 32.60       & $<$4,600   &   $<$0.02   &   $>$29.35$_{+0.30}^{-0.24}$    &   -- \\
Random      &   $0.25 < z \le 0.35$     &   0.332   &   33  &   $>$ 32.85       & $<$9,300   &   $<$0.03   &   $>$29.60$_{+0.33}^{-0.25}$    &   --
\enddata
\tablecomments{Column (5) is the total magnitude corresponding to all of the light within the group. Column (6) represents the equivalent number of stars greater than 3 solar masses (O and B type stars) which would primarily contribute to the UV IGrL (if detected) at the surface brightness limit. Column (7) represents the number of O/B stars per square kpc - assuming the median R$_{gp}$ of the groups in the stack. Column (8) is in units of \rm{mag\,arcsec}$^{-2}$. We utilize a bootstrap analysis with replacement for the uncertainty in the SB upper limit.}
\end{deluxetable*}
\normalsize

\begin{deluxetable*}{lccccccc}[t]
\tabletypesize{\scriptsize}
\tablecolumns{7}
\tablecaption{Physical Group Stack Measurements}
\label{tab:physical_stack}
\tablehead{
\colhead{Stack Type}        &       \colhead{Redshift Range}    &   \colhead{$z_{\rm{med}}$} &       \colhead{Number Groups} &       \colhead{Mag$_{50kpc}$}     &       \colhead{Mag$_{100kpc}$}   & \colhead{$\mu ^{AB}_{U}$}  & \colhead{Excess over Random}  \\
\colhead{(1)}                    & \colhead{(2)}             & \colhead{(3)}                 &  \colhead{(4)}                    & \colhead{(5)}    &     \colhead{(6)}  &     \colhead{(7)}      & \colhead{(8)}                  
}
\startdata
Observed        &   $0.1 < z \le 0.2$       &   0.166   &   17  &   $>$ 32.37   &   $>$ 32.39   &   $>$29.12$_{+0.37}^{-0.28}$    &    -28.2\%      \\
Observed        &   $0.15 < z \le 0.25$     &   0.220   &   27  &   $>$ 32.26   &   $>$ 32.27   &   $>$29.00$_{+0.39}^{-0.29}$    &    27.1\%      \\
Observed        &   $0.25 < z \le 0.35$     &   0.332   &   33  &   $>$ 32.57   &   $>$ 32.58   &   $>$29.32$_{+0.26}^{-0.21}$    &    25.9\%     \\
\hline
Random          &   $0.1 < z \le 0.2$       &   0.166   &   17  &   $>$ 32.01   &   $>$ 32.00   &   $>$28.76$_{+0.63}^{-0.39}$    &    --    \\
Random          &   $0.15 < z \le 0.25$     &   0.220   &   27  &   $>$ 32.50   &   $>$ 32.47   &   $>$29.26$_{+0.40}^{-0.29}$    &    --      \\
Random          &   $0.25 < z \le 0.35$     &   0.332   &   33  &   $>$ 32.82   &   $>$ 32.79   &   $>$29.57$_{+0.31}^{-0.24}$    &     --  
\enddata
\tablecomments{Columns (5) and (6) are the total magnitudes corresponding to all of the light within the central 50 or 100\,kpc. Column (7) is in units of \rm{mag\,arcsec}$^{-2}$. We utilize a bootstrap analysis with replacement for the uncertainty in the SB upper limit. }
\end{deluxetable*}
\normalsize

Placing this amount of light into a different context, we can determine approximately how many O and B type stars are required to account for this UV light. By using the median redshift of the stack, the fraction of the blackbody flux from O/B stars ($>3\,M_\odot$) observed by the LBT U$_{spec}$ filter was determined. Then, by calculating the total IGrL luminosity, the number of O/B stars able to contribute towards the group light can be calculated assuming a Salpeter initial mass function \citep{Salpeter1955}. We find that at most, $\lesssim$3,100 O/B type stars (or $\sim$0.02 O/B stars per kpc$^2$ across the group's projected area) are sufficient to account for the IGrL upper limits for the low redshift group stack, which indicates that minimal star formation is occurring outside of member galaxies in groups. 

The physical stacking method allows for the amount of IGrL to be determined within the central 50 and 100\,kpc of each group. We find no significant difference between the IGrL upper limit within the inner 50\,kpc vs the inner 100\,kpc. This either indicates that there is no additional IGrL outside of the central 50\,kpc of groups on average, or that an IGrL profile may exist but at levels too faint to be detected with these stacks.

\section{Discussion and Summary} \label{sec:discussion}
Here, we summarize and discuss the main results of this study:
\begin{enumerate}
\item We used 25 partial nights with the LBT/LBC between 2007 and 2020 to observe the COSMOS field in the \Ub and create optimal resolution and optimal depth mosaics totaling up to $\sim$37\,hours. Following the seeing sorted stacking process detailed in \citet{Ashcraft2018}, \citet{Otteson2021}, \citet{Redshaw2022} and \citet{Ashcraft2023}, the seeing of each individual exposure was calculated through the median FWHM of $\sim$100 unsaturated stars. Prior to creating the optimal resolution (FWHM $\le 0\farcs9$) and optimal depth (FWHM $\le 1\farcs9$) mosaics, the relative atmospheric transparency was corrected for each exposure. \citet{Ashcraft2018} noted a $\sim$0.2\,mag difference in photometric zero point between the GOODS-N optimal resolution and depth mosaics, which was attributed to variable atmospheric conditions on Mt. Graham \citep{Taylor2004}. However, by correcting for the relative atmospheric transparency differences from exposure to exposure, the difference in photometric zero point was reduced to $\sim$0.05\,mag. The optimal resolution and optimal depth mosaics reached 3$\sigma$ depths of $\sim$26 and $\sim$26.5\,mag, respectively, which are comparable to other U-band surveys \citep{Ashcraft2018, Sawicki2019, Otteson2021,Redshaw2022}

\item Using the optimal depth mosaic in addition to the \citet{Knobel2012} zCOSMOS 20k group catalog, we searched for signatures of IGrL by masking out known sources and only stacking the group backgrounds. We created stacks over three redshift ranges ($0.1 < z \le 0.2$ (N=17), $0.15 < z \le 0.25$ (N=27) and $0.25 < z \le 0.35$ (N=33)) in an effort to trace diffuse UV light in the group environment using two different stacking methods. The fractional stacking method rescaled each group such that each group's radius extended across the same number of pixels, which allowed the IGrL to be probed as a function of group radius. On the other hand, the physical stacking method rescaled each group such that each pixel represented the same size (kpc/pix) across each group.

\begin{figure*}
\centering
\plotone{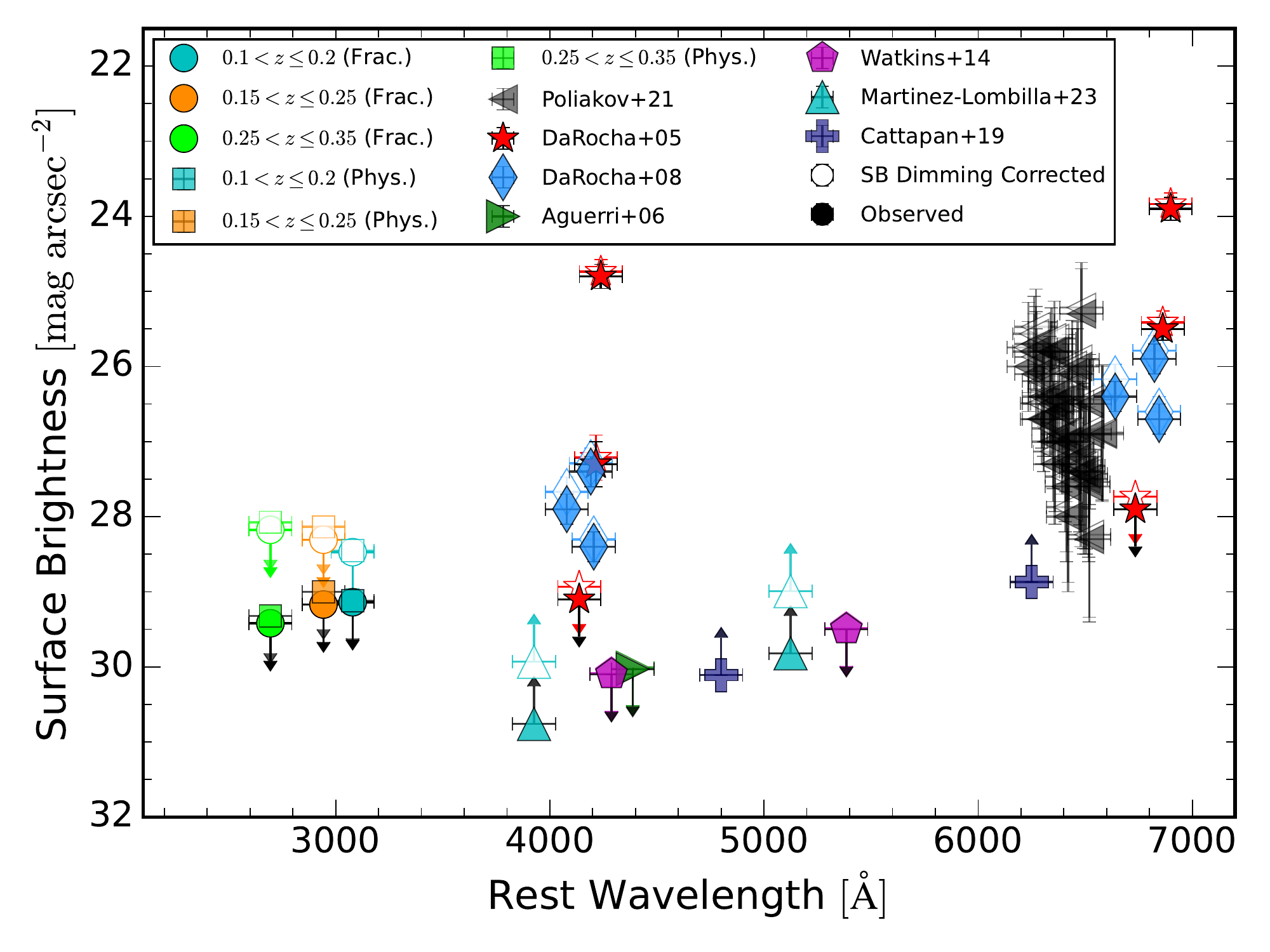}
\caption{IGrL surface brightness measurements as a function of rest frame wavelength for the stacks presented in this work as well as other studies of HCGs \citep{DaRocha2005, Aguerri2006, DaRocha2008, Poliakov2021}, the Leo group \citep{Watkins2014}, and other loose groups \citep{Cattapan2019,Martinez2023}. The points corresponding to \citet{Cattapan2019} and \citet{Martinez2023} were drawn as lower limits to denote that their measurements represented the limiting surface brightness of the outer regions of their IGrL detections. The filled in data points represent the observed IGrL surface brightness, while the outlined points are corrected for the effects of surface brightness dimming. On average, the IGrL detection rate appears to decline as rest frame wavelength shifts bluer, which indicates that the IGrL may not exhibit a strong contribution from young stars. }
\label{fig:sb_comp}
\end{figure*}

\item We find 3$\sigma$ upper limits for the amount of UV IGrL down to $\sim$29.1--29.6\,mag\,arcsec$^{-2}$ for each group stack, which corresponds to $\lesssim 1\%$ of the total group light. Figure \ref{fig:sb_comp} shows the IGrL upper limits measured in this study compared to published values of HCGs \citep{DaRocha2005, Aguerri2006, DaRocha2008, Poliakov2021} and from the Leo group \citep{Watkins2014}. In particular, each of our upper limits are relatively consistent with the IGrL upper limits in HGCs 44 \citep{Aguerri2006}, 88 \citep{DaRocha2005}, and the Leo group \citep{Watkins2014}, albeit at higher redshift and in the rest frame UV. On average, the IGrL appears brighter at longer wavelengths, which may indicate that young stars do not contribute significantly to the IGrL and that minimal in-situ star formation occurs in the IGrL. This would favor tidal stripping and galaxy interactions as the primary method for the formation and buildup of IGrL as proposed by \citet{DaRocha2005, Aguerri2006} and \citet{Poliakov2021} for compact groups and \citet{Spavone2018,Cattapan2019,Raj2020} and \citet{Martinez2023} for loose groups.

\item Despite past observations of the presence of significant amounts of atomic gas in the IGrM \citep{VerdesMontenegro2001,Borthakur2010,Pisano2011,Borthakur2015,Cluver2016,Borthakur2019,Dzudzar2019,McCabe2021,Roychowdhury2022}, we do not observe strong signatures of recent star formation. The lack of UV IGrL detected in this study suggests that the atomic gas structures in the IGrM are likely not dense enough to trigger large scale star formation.

\end{enumerate}

Lastly, we discuss our main findings in context of other published works. Previous studies of IGrL in HCGs by \citet{DaRocha2005}, \citet{Aguerri2006} and \citet{Poliakov2021} constructed a picture where galaxy interactions and tidal stripping remove stars and gas from group members, which eventually settle towards the dominant gravitational potential. In this model, only the most dynamically evolved groups have experienced enough crossing times and interactions necessary for the build up of IGrL. This hypothesis is reinforced by studies of loose groups where IGrL detections were observed to be a direct result of gravitational interactions between group members \citep{Spavone2018,Cattapan2019,Raj2020,Martinez2023}.

In addition, new models and observations by \citet{Kolcu2022} show that, while not as common as in cluster environments, ram pressure stripping is still prevalent in groups. This process was determined to be most consistent with recent infalling galaxies, which may remove cold gas from the infalling member. If this stripped gas is able to fuel star formation, there may be a pathway for IGrL to have a strong contribution from young stars. However, the lack of a UV IGrL detection in our analysis favors tidal interactions over ram pressure stripping as the dominant means for the build up of IGrL. 

The role of active galactic nuclei (AGN) in group environments could also drastically influence the amount of IGrL. \citet{Cui2014} used simulated galaxy clusters and found evidence that AGN activity is able to alter the fraction of diffuse, stellar light by up to a factor of 2. AGN feedback may prevent star formation in the central regions of galaxy groups due to the sheer amount of energy returned to the IGrM \citep[and references therein]{McNamara2007,Gitti2012,Heckman2014,Gaspari2020,Eckert2021}. If this is the case, it might explain the lack of star formation in the IGrM and the corresponding faint limits to the U-band IGrL in this study. 

Comparing these IGrL properties to intracluster light (ICL), we find similarities in terms of possible formation processes. The main theories for the buildup of ICL are any of the following (or a combination thereof): tidal stripping \citep{Rudick2009}, the ejection of stars from galaxy mergers \citep{Willman2004,Murante2007}, and the accretion of galaxy groups \citep{Mihos2004, Rudick2006}. However, \citet[and references therein]{Montes2022} note that studying the ICL through stellar populations may only unveil the dominant formation process of a particular cluster and not a combination of mechanisms which may vary from cluster to cluster. For the Virgo and Coma clusters, \citet{Williams2007} and \citet[and references therein]{Coccato2010} find that the ICL is primarily composed of older stars, which align with our conclusions of minimal star formation in groups that could contribute to IGrL. However, \citet{Jimenez2016} find an increased amount of ICL in the rest frame B-band for Abell 2744, which is indicative of a younger population of stars that may have formed during merging events. Given the large amounts of atomic gas known to exist in groups \citep{VerdesMontenegro2001,Borthakur2010,Pisano2011,Borthakur2015,Cluver2016,Borthakur2019,Dzudzar2019,McCabe2021,Roychowdhury2022}, merging groups may be a potential environment to search for star formation in the IGrM traced by U-band IGrL. 

Future studies may be able to determine the dependence of IGrL on both wavelength and dynamical evolution. The majority of the studies shown in Figure \ref{fig:sb_comp} represent studies of HCGs, which are ideal environments for galaxy interactions. Observations of a wider range of group environments could provide insight into the build up of IGrL through galaxy interactions and/or tidal stripping. In \cite{Ashcraft2023}, evidence of additional diffuse light in the outskirts of galaxies was detected in the $r$-band resulting from tidal tails and galaxy interactions. These signs of interactions have also been abundantly observed with JWST \citep[and references therein]{Finkelstein2022,Windhorst2023}. Additionally, filling in the wavelength gaps in Figure \ref{fig:sb_comp} may help place constraints on the stellar populations contributing to IGrL. 

Finally, increasing the number of publicly available galaxy group catalogs in popular survey fields would be beneficial to the study of IGrL. Similar group stacks could be created in other survey fields where deep, multi-wavelength data already exists. Moreover, more group catalogs would aid in other studies using QSO absorption lines to probe diffuse gas in group environments. The combination of IGrM studies in absorption and emission via IGrL are necessary in order to fully understand the impact that group environments have on galaxy evolution.

\begin{acknowledgments}
The Arizona State University authors acknowledge the twenty three Native Nations that have inhabited this land for centuries. Arizona State University’s four campuses are located in the Salt River Valley on ancestral territories of Indigenous peoples, including the Akimel O’odham (Pima) and Pee Posh (Maricopa) Indian Communities, whose care and keeping of these lands allows us to be here today. We acknowledge the sovereignty of these nations and seek to foster an environment of success and possibility for Native American students and patrons.

T.M. would like to thank Brad Koplitz, Mansi Padave, and Alejandro Olvera for their thoughtful discussion and input throughout this project. The authors acknowledge support from UVCANDELS grant HST-GO-15647 provided by NASA through the Space Telescope Science Institute, which is operated by the Association of Universities for Research in Astronomy, Inc., under NASA contract NAS 5-2655, and from NASA JWST Interdisciplinary Scientist grants NAG5-12460, NNX14AN10G and 80NSSC18K0200 from GSFC.

The LBT is an international collaboration among institutions in the United States, Italy, and Germany. LBT Corporation partners are The University of Arizona on behalf of the Arizona university system; Istituto Nazionale di Astrofisica, Italy; LBT Beteiligungsgesellschaft, Germany, representing the Max-Planck Society, the Astrophysical Institute Potsdam, and Heidelberg University; The Ohio State University; and The Research Corporation, on behalf of The University of Notre Dame, University of Minnesota, and University of Virginia. 

\software{Astropy \citep{Astropy2013,Astropy2018,Astropy2022}}
\software{SExtractor \citep{Bertin1996}}
\software{Swarp \citep{Bertin2002,Bertin2010}}

\end{acknowledgments}

\newpage
\vspace{2cm}
\bibliography{References}
\bibliographystyle{aasjournal}


\end{document}